%% file: lanz.tex
\documentclass[conference, 10pt]{IEEEtran}

\usepackage{graphicx}
\usepackage{color}
\usepackage{placeins}
\usepackage{float}
\usepackage{tabularx,colortbl}
\usepackage{times,amsmath}
\usepackage{amsmath,graphicx}
\usepackage{amssymb}

\usepackage{subfig}
\usepackage{hhline}

\usepackage{psfrag}
\usepackage[table,xcdraw,dvipsnames]{xcolor}
\usepackage{setspace}

\usepackage{placeins}

\usepackage{pgfplots}
\pgfplotsset{compat=newest}
\pgfplotsset{plot coordinates/math parser=false} 
\usepgfplotslibrary{groupplots}
\usepackage{tikz}
\usetikzlibrary{matrix}
\newlength\figureheight
\newlength\figurewidth 

\begin{document}
%
\title{Compression of Dynamic Medical CT Data Using Motion Compensated Wavelet Lifting with Denoised Update}

\author{\IEEEauthorblockN{Daniela Lanz, J\"{u}rgen Seiler, Karina Jaskolka, and Andr\'{e} Kaup}
\IEEEauthorblockA{Multimedia Communications and Signal Processing\\
Friedrich-Alexander-Universit\"at Erlangen-N\"urnberg (FAU)\\
Cauerstr. 7, 91058 Erlangen, Germany\\
Email: {Daniela.Lanz,Juergen.Seiler,Karina.Jaskolka,Andre.Kaup}@FAU.de\\}}

\maketitle

\begin{abstract}
\boldmath
For the lossless compression of dynamic \mbox{3-D+t} volumes as produced by medical devices like Computed Tomography, various coding schemes can be applied. This paper shows that \mbox{3-D} subband coding outperforms lossless HEVC coding and additionally provides a scalable representation, which is often required in telemedicine applications. However, the resulting lowpass subband, which shall be used as a downscaled representative of the whole original sequence, contains a lot of ghosting artifacts. This can be alleviated by incorporating motion compensation methods into the subband coder. This results in a high quality lowpass subband but also leads to a lower compression ratio. In order to cope with this, we introduce a new approach for improving the compression efficiency of compensated 3-D wavelet lifting by performing denoising in the update step. We are able to reduce the file size of the lowpass subband by up to $1.64\%$, while the lowpass subband is still applicable for being used as a downscaled representative of the whole original sequence.
\end{abstract}

\IEEEpeerreviewmaketitle

\section{Introduction}
\label{sec:intro}

\begin{figure}[t]
	\begin{center}
		\psfragscanon
		\psfrag{x}{$x$}
		\psfrag{y}{$y$}
		\psfrag{z}{$z$}
		\psfrag{t}{$t$}
		\psfrag{X}{$X$}
		\psfrag{Y}{$Y$}
		\psfrag{Z}{$Z$}
		\psfrag{T}{$T$}
		\psfrag{1}{$1$}
		\psfrag{2}{$2$}
		\includegraphics[width=0.48\textwidth]{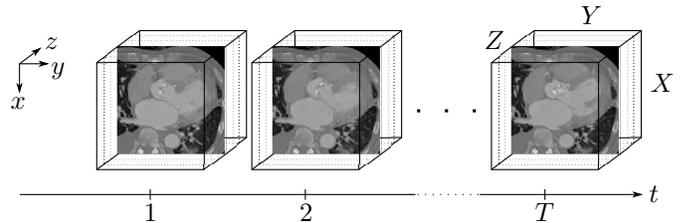}
		\psfragscanoff
		\caption{Example of a dynamic medical data set: The sketch shows a 3-D+t CT volume of a thorax consisting of subsequent 3-D volumes over time $t$.}
		\label{fig:data}
		\vspace{-14pt}
	\end{center}
\end{figure}

In the daily medical routine, dynamic volume data provide a good basis for analyses and predictions of spatio-temporal movements of particular parts of the human body. Fig.~\ref{fig:data} shows the principal structure of a \mbox{3-D+t} volume from Computed Tomography (CT). It consists of $T$ temporally equidistant \mbox{3-D} volumes of size \mbox{$X$$\times$$Y$$\times$$Z$}. Due to the high temporal and spatial resolution, dynamic volumes can get very large, which makes storing and archiving them in a lossless manner impractical. Additionally, for telemedicine applications a scalable representation is often required that allows for tasks like browsing and fast previewing~\cite{Doukas:2008:ATM:1556147.1362837}. 
Moreover, CT data contain a lot of sensor noise. This is caused by the radiation which has to be kept low to reduce the risks for the patients as well as the short acquisition time which is kept as low as possible to avoid motion artifacts. 
Further, medical CT data have a higher bit depth than natural video sequences, namely 12 bits per pixel. 
Therefore, an efficient coding scheme is required.

Common video codecs are mainly designed for temporal 8-bit video sequences originating from the entertainment industry. 
One way to apply them on medical 12-bit \mbox{3-D+t} data is to generate a single sequence over $t$ for every slice position $z$, resulting in $Z$ temporal sequences. In Fig.~\ref{fig:data}, the resulting sequence for slice position $z=3$ is highlighted. Each temporal sequence can be compressed using common video coding schemes under the condition of lossless compression and adaption of the range of the bit depth.

The HEVC standard~\cite{2016}, which describes a motion compensated predictive coder, is mainly applied for the efficient coding of video sequences. By choosing the \textit{Lowdelay Main RExt} configuration and lossless mode, it is possible to apply HEVC also to medical sequences. An alternative coding scheme to predictive coding is represented by 3-D subband coding (SBC). Incorporating motion compensation (MC) methods into the subband coder is called Motion Compensated Temporal Filtering~(MCTF)~\cite{334985}. 

In~\cite{8066388}, dynamic volume data is compressed by performing one Haar wavelet transform (WT) in temporal direction. This WT is performed in a simple SBC coder and in an MCTF coder with mesh-based MC. In both cases, the resulting subbands are coded losslessly using the wavelet-based volume coder JPEG~2000~\cite{ITU-T-800} with four spatial decompositions steps.

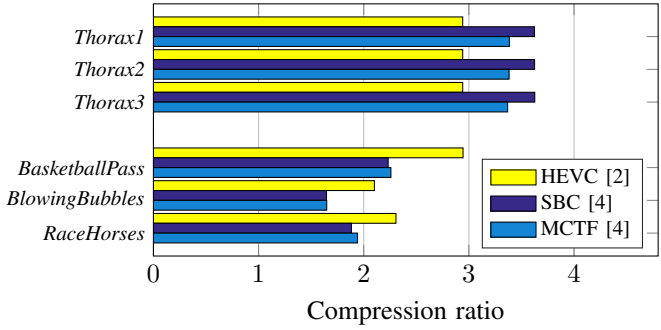
\begin{figure}[t]
	\begin{center}
		\setlength\figurewidth{0.37\textwidth}
		\input{Figures/hevc.tikz}
		\caption{Compression ratio resulting from HEVC, SBC, and mesh-based MCTF for medical (top) as well as natural sequences (bottom).}
		\label{fig:hevc}
	\end{center}
\end{figure}

Fig.~\ref{fig:hevc} shows the resulting compression ratios of the above mentioned coders for medical as well as natural sequences. For the HEVC codec, the latest test model HM-16.16 was used.
The medical sequences \textit{Thorax1-3} originate from a \mbox{3-D+t} CT data set at slice positions $z=1,2,3$. The content of this volume can exemplarily be seen in Fig.~\ref{fig:data} and describes a beating heart. The natural sequences consist of three HEVC-specific \textit{ClassD} sequences~\cite{Bossen2013}.
As the medical sequences comprise luminance information only, all sequences are used in 4:0:0 color sub-sampling format for a fair comparison.

As can be seen in Fig.~\ref{fig:hevc}, HEVC reaches good compression ratios for natural sequences, but performs relatively less efficient for CT data. Apart from this, HEVC offers no scalable representation for the original volume.
In contrast, SBC performs better than HEVC on medical CT data and additionally provides scalability features.
However, SBC causes ghosting artifacts in the lowpass (LP) subband due to temporal displacements in the sequence. Thus, SBC is not recommended if the LP subband is to be used as a downscaled representative of the whole original sequence.
MCTF performs not as well as SBC regarding the compression efficiency of medical CT data, but provides a high quality LP subband. Accordingly, MCTF results in a proper scalable representation for telemedicine applications.

To improve the compression efficiency of MCTF, we propose to apply denoising in the update step. With this, we avoid warping noise from the highpass (HP) subband to the LP subband. This novel step in the context of compensated wavelet lifting leads to a lower entropy in the LP subband and the compression ratio can be increased. By applying adequate filters for denoising, the compression ratio rises while the suitability of the LP subband for telemedicine applications is preserved at the same time.


After a short recap of compensated wavelet lifting in Section~\ref{sec:WT}, Section~\ref{sec:denoising} provides a detailed description of the proposed method, followed by the simulation results in Section~\ref{sec:results}. In Section~\ref{sec:conclusion} we give a short conclusion and outlook.

\section{Compensated Wavelet Lifting}
\label{sec:WT}

\begin{figure}[t]
	\centering
	\psfragscanon
	\psfrag{f2t}{$f_{2t}$}
	\psfrag{f2t-1}{$f_{2t-1}$}
	\psfrag{HP}{$\text{HP}_t$}
	\psfrag{LP}{$\text{LP}_t$}
	\psfrag{MC}{MC}
	\psfrag{IMC}{$\text{MC}^{-1}$}
	\psfrag{B}{\textcolor{red}{DN}}
	\psfragscanoff
	\includegraphics[width=0.48\textwidth]{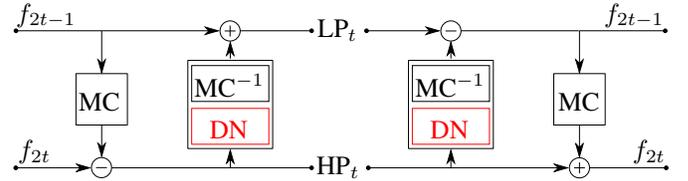}
	\caption{Block diagram of the lifting structure containing the proposed denoising (DN) filter at both encoder (left) and decoder side (right).}
	\label{fig:lifting_filtered}
\end{figure}

An efficient implementation of the discrete WT, named lifting structure, was proposed by Swelden~\cite{Sweldens1995}. The first step is a decomposition of the input video signal into even- and odd-indexed frames $f_{2t}$ and $f_{2t-1}$. In a prediction step, the odd frames are predicted and subtracted from the even frames, resulting in the HP frames. Then, in an update step, the HP frames are filtered and added back to the odd frames, resulting in the LP frames. 
Fig.~\ref{fig:lifting_filtered} shows a block diagram of the lifting structure. The temporal HP and LP frames are generated by
\begin{align}
\text{HP}_t &= f_{2t}-P(f_{2t-1}) \label{eq:HP}\\
\text{LP}_t &= f_{2t-1}+U(\text{HP}_t) \label{eq:LP},
\end{align}
where $P(\cdot)$ and $U(\cdot)$ describe the prediction and update operators respectively. 
In the prediction step, MC is usually done to avoid ghosting artifacts in the LP frames that are caused by temporal displacements in the sequence~\cite{958672}. This MC has to be inverted in the update step.

However, while the appearance of ghosting artifacts is reduced by MC, the noise variance of the single subbands is increased. Structural information as well as noise is warped from one frame to the other by the prediction and update operators. As described in~\cite{lnt2012-10}, this leads to a higher overall entropy, so the required number of bits for coding the subbands will also rise.

One possibility to reduce the increase of the noise variance in the LP frames is to skip the update step entirely. This results in the so-called Truncated WT, in which the LP frames are generated by subsampling the sequence by a factor of 2. However, apart from negative effects like temporal aliasing and temporal fluctuation in video quality as discussed in~\cite{1246794}, the LP subband is not suitable anymore for telemedicine applications. An adequate downscaled representative should offer a high similarity to the odd- as well as to the even-indexed frames, which is not given by simply subsampling the original sequence.

\section{Compensated Wavelet Lifting with Denoised Update}
\label{sec:denoising}

To reduce the increase of the noise variance in the LP frames and thereby improving the compression efficiency of MCTF, we propose to apply denoising in the update step as shown in red in Fig.~\ref{fig:lifting_filtered}.
Considering \eqref{eq:HP} and \eqref{eq:LP}, the \underline{W}avelet \underline{L}ifting with \underline{D}enoised \underline{U}pdate (WLDU) is described by
\begin{align}
\text{HP}_t &= f_{2t}-\text{MC}(f_{2t-1}) \label{eq:HP_adapted} \\
\text{LP}_t &= f_{2t-1} + \text{MC}^{-1}(\textcolor{red}{\text{DN}(}\text{HP}_t\textcolor{red}{)}) \label{eq:LP_adapted}.
\end{align}
Since the lifting scheme provides a flexible framework, $f_{2t}$ and $f_{2t-1}$ can be reconstructed without any loss if the denoising filter is also applied at the decoder side, resulting in
\begin{align}
f_{2t-1} &= \text{LP}_t-\text{MC}^{-1}(\textcolor{red}{\text{DN}(}\text{HP}_t\textcolor{red}{)}) \label{eq:odd_adapted} \\
f_{2t} &= \text{HP}_t+ \text{MC}(f_{2t-1}).
\label{eq:even_adapted}
\end{align}
These equations hold for any denoising filter without compromising the property of perfect reconstruction.

Under the assumption that the HP frames are zero-mean, an infinitely strong filter would only add zeros to the odd frames in the update step. This would correspond to the Truncated WT. In theory, the maximum achievable compression ratio of WLDU is hence bounded by the performance of the Truncated WT.
Therefore, we apply a simple 2-D Gaussian filter in a first step so as to verify this \mbox{theoretical} bound of the compression ratio.

After that, we will apply more complex filters. To guarantee an accurate inverse MC in the update step, structural details in the HP frames shall be preserved while noisy structures caused by erroneous motion models and warping processes shall be blurred to avoid augmenting additional noise to the LP frames. 
In the context of video coding, various filters have been used for in-loop denoising of reference frames~\cite{lnt2010-23}, which is why we will also apply them in our novel framework. These filters are the Adaptive Wiener Filter (AWF)~\cite{1990ph...book.....L}, the Non-Local Means algorithm (NLM)~\cite{1467423}, and Block Matching and 3-D Filtering (BM3D)~\cite{4271520}. In addition to these filters, we will apply Guided Image Filtering (GIF)~\cite{He2010}.


All these algorithms are influenced by the filter strength $h$, which can be calculated by 
\begin{equation}
h = \xi \cdot \sigma_{n}^2,
\end{equation}
where $\sigma_{n}^2$ denotes the noise variance of the input image and $\xi$ describes an arbitrary parameter which optimizes the strength of denoising in order to improve the compression efficiency. By increasing $\xi$, the noise variance of the output image is decreased and according to~\cite{lnt2012-10}, a better compression ratio can be reached. 

Noise estimation is done at the encoder side. To guarantee lossless reconstruction, the estimated noise variance $\sigma_{n}^2$ has to be known at the decoder side. This can be assured by transmitting $\sigma_{n}^2$ as side information to the decoder or by estimating $\sigma_{n}^2$ at both the encoder and the decoder side. 
There exist different possibilities to perform noise estimation, which differ mainly with regard to the accuracy and the computational complexity.
To avoid transmitting additional information to the decoder and to keep the decoder-side complexity low, we decided to estimate the noise variance by a low-complexity algorithm proposed in \cite{IMMERKAER1996300} at both the encoder and the decoder side. 

By applying these filters in our novel framework of WLDU, we enforce a higher compression efficiency than MCTF, while the suitability of the LP subband for telemedicine applications is preserved.

\section{Simulation Results}
\label{sec:results}

\begin{figure}[t]
	\centering
	\psfragscanon
	\psfrag{f2t}{$f_{2t}$}
	\psfrag{f2t-1}{$f_{2t-1}$}
	\psfrag{HP}{$\text{HP}_t$}
	\psfrag{LP}{$\text{LP}_t$}
	\psfrag{MC}{MC}
	\psfrag{IMC}{$\text{MC}^{-1}$}
	\psfrag{B}{\textcolor{red}{DN}}
	\psfrag{PSNR}{$\text{PSNR}(f_{2t-1},\text{LP}_t)$}
	\psfrag{P2}{$\text{PSNR}(f_{2t},\text{MC}(\text{LP}_t))$}
	\psfrag{MCLP}{$\text{MC}(\text{LP}_t)$}
	\psfragscanoff
	\includegraphics[width=0.47\textwidth]{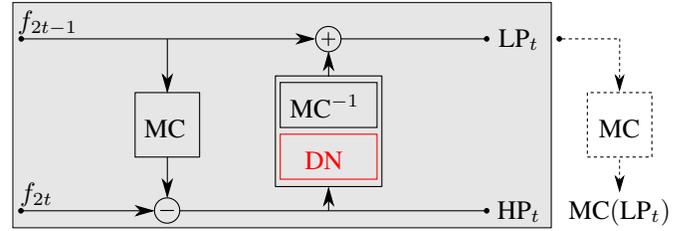}
	\caption{Extended block diagram of the lifting structure illustrating the similarity of $\text{LP}_t$ to $f_{2t}$ and $f_{2t-1}$.}
	\label{fig:metric}
\end{figure}

In our simulation setup, we use a \mbox{3-D+t} medical CT data set\footnote{The CT volume data set was kindly provided by Siemens Healthineers.} that describes a beating heart and comprises 10 time steps, each with 127 slices and a resolution of 512$\times$512 pixels at 12 bits per sample. This results in 127 temporal sequences. The first three sequences correspond to the test sequences \mbox{\textit{Thorax1-3}} in Section~\ref{sec:intro}.
We perform one Haar wavelet decomposition step with a mesh-based MC with and without the proposed denoising step. For the mesh-based MC, we use a grid size of 8$\times$8 pixels. 
The subbands are compressed losslessly using the wavelet-based volume coder JPEG 2000. We use the OpenJPEG~\cite{openjpeg} implementation with four spatial wavelet decomposition steps in $xy$-direction. 

\subsection{Evaluation of the Quality of the Lowpass Subband}
\label{sec:metric}

\begin{figure*}[tb]
	\begin{center}
		\setlength\figurewidth{0.4\textwidth}
		\input{Figures/results_conv_2.tikz}
		\caption{$\text{PSNR}_{\text{LP}_t}$ and $\text{SSIM}_{\text{LP}_t}$ results compared against the file size of the LP frames in [kB]. The arrow above the diagrams shows the direction of the single curves for increasing values of the filter strength $h = \xi\cdot \sigma_{n}^2$. For better presentation, only every 10th value for $h$ is plotted.}
		\label{fig:RD_PSNR}
	\end{center}
\end{figure*}
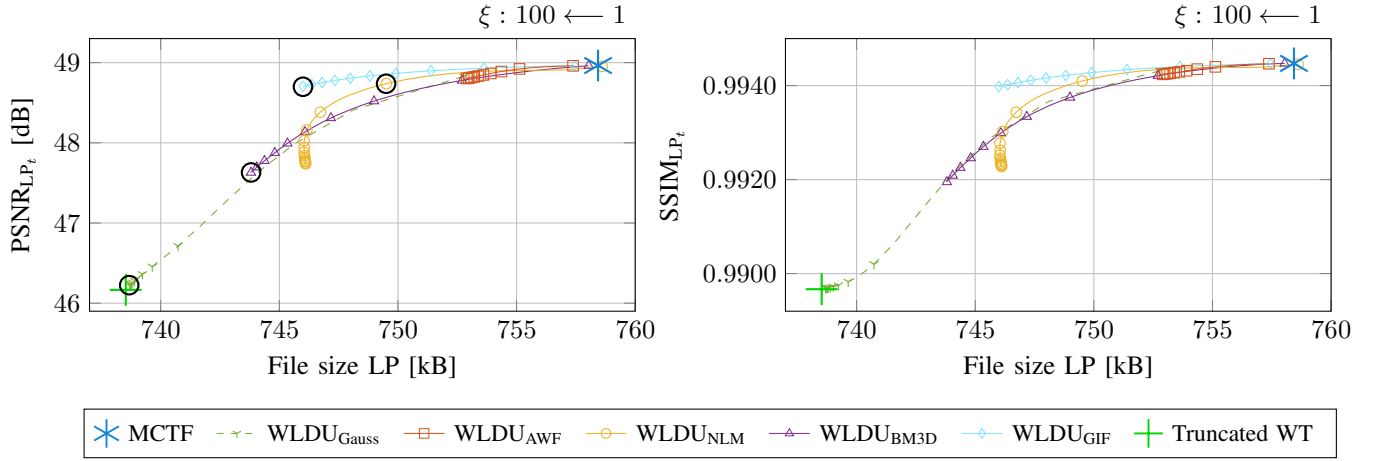

Usually, the quality of the LP subband is measured by evaluating the similarity to the odd-indexed frames in terms of PSNR. However, in many applications the LP subband is to be used as a downscaled representative for the whole original sequence. Therefore, a high similarity between the LP frames and the even-indexed frames $f_{2t}$ should also be considered. As shown in Fig.~\ref{fig:metric}, this can be done by warping each LP frame to the time step of the even-indexed frame and measure their similarity also in terms of PSNR. 
Since PSNR goes to infinity in case of perfect MC, it is not sufficient to calculate the average PSNR. Hence, for evaluating the LP frames of every time step with respect to both mentioned aspects, we suggest two different metrics:
\begin{itemize}
	
	\item First we consider the variance of the error signal consisting of the even- and odd-indexed frames and their corresponding LP frames
	\begin{equation}
		 \sigma_e^2 = \frac{1}{2}\left(\lVert f_{2t-1}-\text{LP}_t\rVert^2 + \lVert f_{2t} - \text{MC}(\text{LP}_t)\rVert^2\right).
	\end{equation}
	Then the quality of each lowpass frame $\text{LP}_t$ can be measured by
	\begin{equation}
	\text{PSNR}_{\text{LP}_t} [\text{dB}]= 10\log_{10}\frac{A_{\text{max}}^2}{\sigma_e^2},
	\end{equation}  
	where $A_{\text{max}}$ corresponds to the maximum possible amplitude of the signal.
	
	\item An alternative to PSNR is given by the Structural Similarity Index (SSIM)~\cite{1284395}. Since SSIM results in a range of [0,1], it is possible to calculate the average value regarding the similarity of each lowpass frame $\text{LP}_t$ to $f_{2t}$ and $f_{2t-1}$:
	\begin{align}
	\begin{split}
	\text{SSIM}_{\text{LP}_t} & = \frac{1}{2}\left( \text{SSIM}(f_{2t-1},\text{LP}_t)+%
	\vphantom{+ \text{SSIM}(f_{2t},\text{MC}(\text{LP}_t))} \right. \\
	& \hphantom{=++}\left. \vphantom{\text{SSIM}(f_{2t-1},\text{LP}_t)}%
	+ \text{SSIM}(f_{2t},\text{MC}(\text{LP}_t)) \right).
	\end{split}
	\end{align}
\end{itemize}
In the following sections, the term ``quality of the lowpass subband'' consequently describes a value calculated by one of these two metrics.

\subsection{Analysis of the Simulation Results}

\begin{table*}[tb]
	\centering
	\caption{File size and quality of the LP subband in terms of $\text{PSNR}_{\text{LP}_t}$ for certain values of Fig.~\ref{fig:RD_PSNR}. Absolute and relative differences against MCTF are also provided. The line printed in bold indicates the setup that we recommend for the given data set.}
	\label{tab:filesizes}
	\begin{tabular}{l|c|c|c||c|c}
		&File size LP {[}kB{]} & \multicolumn{2}{c||}{$\Delta$ to MCTF} & $\text{PSNR}_{\text{LP}_t}$ [dB] & $\Delta$ to MCTF [dB]\\ \hhline{~|~|--||~|~}  
		&      & absolute [kB] & relative $\left[\%\right]$ &      & \\ \hline
		MCTF & 758.43 & -      & -      & 48.96 &  -   \\
		$\text{WLDU}_\text{NLM}$	& 749.51 & - 8.92 & -1.18 & 48.74 & -0.22 \\
		\textbf{$\text{WLDU}_\text{GIF}$}	& \textbf{745.99} & \textbf{-12.44} & \textbf{-1.64} & \textbf{48.71} & \textbf{-0.25} \\
		$\text{WLDU}_\text{BM3D}$	& 743.80 & -14.63 & -1.92 & 47.63 & -1.33 \\
		$\text{WLDU}_\text{Gauss}$	& 738.69 & -19.74 & -2.60 & 46.22 & -2.74 \\
		Truncated WT				& 738.53 & -19.90 & -2.62 & 46.17 & -2.79 
	\end{tabular}
\end{table*}


Fig.~\ref{fig:RD_PSNR} shows the $\text{PSNR}_{\text{LP}_t}$ and $\text{SSIM}_{\text{LP}_t}$ results over the file size of the LP subband in [kB] for all considered filter setups.
We examine the influence of the filter strength $h$ by varying $\xi$ in a range of integer values from $1$ to $100$. The results are averaged over all frames of all sequences.

As mentioned in Section~\ref{sec:denoising}, we verify the upper bound of the maximum achievable compression ratio by applying a simple \mbox{2-D} Gaussian filter. 
The dashed green curve in Fig.~\ref{fig:RD_PSNR} shows the development of $\text{WLDU}_\text{Gauss}$ for increasing values of $h$. For high filter strengths the results are very close to the performance of the Truncated WT. However, the quality of the LP subband in terms of $\text{PSNR}_{\text{LP}_t}$ as well as $\text{SSIM}_{\text{LP}_t}$ decays rapidly, making it useless for telemedicine applications.

Any curve which lies above the curve of $\text{WLDU}_\text{Gauss}$ results in a LP subband with a higher quality calculated by $\text{PSNR}_{\text{LP}_t}$ or $\text{SSIM}_{\text{LP}_t}$ that may be used in telemedicine applications. 
We are looking for a filter which keeps the quality of the LP subband at the level of MCTF for as long as possible. It should not decay until the upper bound for the compression ratio is almost achieved.

Therefore, we apply more complex denoising techniques as introduced in Section~\ref{sec:denoising}. For the AWF, we use a window of $3$$\times$$3$ pixels. For the NLM algorithm, the support area has a size of $5$$\times$$5$ pixels and the neighborhood size is $3$$\times$$3$ pixels. The implementation used for BM3D is provided by~\cite{Dabov2007}. 
GIF is applied under self-guidance, using $\text{HP}_t$ itself as the guidance image. The window used in GIF is of size $5$$\times$$5$ pixels.

According to Fig.~\ref{fig:RD_PSNR}, AWF seems not to be the right choice in the context of WLDU, since no gain regarding neither the compression ratio nor the quality calculated by $\text{PSNR}_{\text{LP}_t}$ or $\text{SSIM}_{\text{LP}_t}$ can be reached compared to $\text{WLDU}_\text{Gauss}$. 
$\text{WLDU}_\text{BM3D}$ has a quite high computational complexity but gives no significant advantage compared to $\text{WLDU}_\text{Gauss}$. Therefore, it is also not suited for WLDU.

In contrast, by applying GIF and NLM as filters in the context of WLDU, we are able to reach higher compression ratios at nearly constant quality of the LP subband for small values of $h$. However, for higher filter strengths, $\text{WLDU}_\text{NLM}$ completely fails in terms of $\text{PSNR}_{\text{LP}_t}$ as well as $\text{SSIM}_{\text{LP}_t}$. In contrast, $\text{WLDU}_\text{GIF}$ keeps the quality of the LP subband at a high level even for high filter strengths. With $\text{WLDU}_\text{GIF}$, we thus found a filter which fulfills the desired behavior: We achieve a higher compression efficiency of the LP subband at a quality close to MCTF with regard to both metrics $\text{PSNR}_{\text{LP}_t}$ and $\text{SSIM}_{\text{LP}_t}$. 

For a closer examination, we choose one value in Fig.~\ref{fig:RD_PSNR} that is good in a rate-distortion sense for each $\text{WLDU}_\text{NLM}$, $\text{WLDU}_\text{BM3D}$, and $\text{WLDU}_\text{GIF}$. Additionally, we choose the lowest possible value which we can achieve by WLDU regarding the file size of the LP subband. This value belongs to $\text{WLDU}_\text{Gauss}$. All these values are marked with black circles in Fig.~\ref{fig:RD_PSNR}.
The corresponding values for the quality in terms of $\text{PSNR}_{\text{LP}_t}$ and the file size of the LP subband can be found in Table~\ref{tab:filesizes}. 

By applying $\text{WLDU}_\text{Gauss}$, we can save $19.74$~kB compared to MCTF, which corresponds to $2.60\%$ bit rate savings. However, $\text{PSNR}_{\text{LP}_t}$ amounts only to $46.22$~dB. This loss of $2.74$~dB constitutes the inability of the resulting LP subband for being used in telemedicine applications.
In contrast, by applying the more complex filters, the quality of the LP subband in terms of $\text{PSNR}_{\text{LP}_t}$ is significantly less degraded, while the file size can still be reduced by more than $1\%$.
In particular, choosing $\text{WLDU}_\text{GIF}$ saves  $12.44$~kB compared to MCTF. This corresponds to bit rate savings of $1.64\%$ at a loss of only $0.25$~dB regarding the quality of the LP subband. Consequently, the applicability of the LP subband to represent the whole original volume is preserved and the compression efficiency is increased at the same time.

For very high filter strengths, however, the results stagnate for all applied filters. The theoretical upper bound of the maximum achievable compression ratio given by the Truncated WT cannot be reached by further increasing $\xi$. 
With larger neighborhood sizes for the single filters, a further compression would be possible. However, this would result in significantly lower values for $\text{PSNR}_{\text{LP}_t}$ as well as for $\text{SSIM}_{\text{LP}_t}$, which is not useful, if the LP subband is to be used as a downscaled representative.


\section{Conclusion}
\label{sec:conclusion}

In this paper, a novel technique for improving the compression efficiency of MCTF at nearly constant quality of the LP subband in terms of $\text{PSNR}_{\text{LP}_t}$ as well as $\text{SSIM}_{\text{LP}_t}$ was proposed. 
After demonstrating that HEVC is not as efficient as SBC with regard to the compression of dynamic CT data, it was shown that SBC gives no satisfying scalable representation for use in telemedicine applications. Incorporating MC methods into the lifting structure of SBC was shown to result in a high quality LP subband, while suffering with regard to the compression efficiency. Therefore, we proposed to apply denoising in the update step, called WLDU.
This novel approach preserves the suitability of the LP subband to be used as a downscaled representative of the whole original sequence and improves the compression efficiency at the same time.
Further work aims at the investigation of optimum denoising filters and the suitability of deblocking filters for block-based MC.
\section*{ACKNOWLEDGEMENT}
We gratefully acknowledge that this work has been supported by the Deutsche Forschungsgemeinschaft (DFG) under contract number KA 926/4-3.

\bibliographystyle{IEEE}
\bibliography{Literatur}

\end{document}

%% file: Figures/hevc.tikz
%
%
\definecolor{mycolor1}{rgb}{0.20810,0.16630,0.52920}%
\definecolor{mycolor2}{rgb}{0.07935,0.52002,0.83118}%
\definecolor{mycolor4}{rgb}{1,1,0}%
\begin{tikzpicture}

\begin{axis}[%
width=\figurewidth ,
height=0.5\figurewidth ,
at={(0\figurewidth ,0\figurewidth )},
scale only axis,
xmajorgrids,
y dir=reverse,
ymin=0,
ymax=7.7,
ytick={1,2,3,5,6,7},
yticklabel style = {font=\footnotesize,, inner sep=1mm},
yticklabels={\textit{{Thorax1}},\textit{{Thorax2}},\textit{{Thorax3}},\textit{{BasketballPass}},\textit{{BlowingBubbles}},\textit{{RaceHorses}}},
xmin=0,
xmax=4.8,
xlabel={Compression ratio},
axis background/.style={fill=white},
legend pos={south east},
legend style={legend cell align=left,align=left,nodes={scale=0.8},draw=none,draw=white!15!black}
]
\addplot[xbar,bar width = 0.3, bar shift = -0.45, draw=black,fill=mycolor4,area legend] plot table[row sep=crcr] {%
	2.941176471 1\\
	2.941176471	2\\
	2.941176471	3\\
	0			4\\
	2.94444444	5\\
	2.101321586	6\\
	2.306451613	7\\	
};
\addlegendentry{HEVC \cite{2016}};

\addplot[xbar,bar width=0.3,bar shift=-0.15,draw=black,fill=mycolor1,area legend] plot table[row sep=crcr] {%
3.6234	1\\
3.6248	2\\
3.6265	3\\
0		4\\
2.232048852	5\\
1.645957053	6\\
1.882569774	7\\
};
\addlegendentry{SBC \cite{8066388}};


\addplot[xbar,bar width=0.3, bar shift= 0.15,draw=black,fill=mycolor2,area legend] plot table[row sep=crcr] {%
3.3853	1\\
3.3823	2\\
3.3685	3\\
0		4\\
2.258287489	5\\
1.647879004	6\\
1.940812019	7\\
};
\addlegendentry{MCTF \cite{8066388}};

\end{axis}
\end{tikzpicture}%

%% file: Figures/results_conv_2.tikz
%
%
\definecolor{mycolor1}{rgb}{0.07935,0.52002,0.83118}%
\definecolor{mycolor2}{rgb}{0.85000,0.32500,0.09800}%
\definecolor{mycolor3}{rgb}{0.92900,0.69400,0.12500}%
\definecolor{mycolor4}{rgb}{0.49400,0.18400,0.55600}%
\definecolor{mycolor5}{rgb}{0.502,0.8980,1}%
\definecolor{mycolor6}{rgb}{0,0.8313,0}%
\definecolor{mycolor7}{rgb}{0.46600,0.67400,0.18800}%
\begin{tikzpicture}

\begin{groupplot}[group style={group name = group,group size=2 by 1, horizontal sep = 2cm, vertical sep=1.1cm}]

\nextgroupplot[%
width=\figurewidth ,
height=0.5\figurewidth ,
scale only axis,
xmin=737,
xmax=760,
xlabel={File size LP [kB]},
xmajorgrids,
ymin=45.9,
ymax=49.3,
ylabel={$\text{PSNR}_{\text{LP}_t}$ [dB]},
title={$\xi:$ $100\longleftarrow 1$},
title style={at={(1,1.1)},anchor=north east},
ymajorgrids,
axis background/.style={fill=white},
]
\addplot [color=mycolor7,dashed,mark=Mercedes star flipped,mark options={solid},mark repeat = 11,mark phase = 1]
table[row sep=crcr]{%
752.783372293307	48.8303628608179\\
748.09842519685	48.4078109684273\\
745.649514025591	47.9892615021332\\
744.201417937992	47.6581645842479\\
743.273560531496	47.4101735817014\\
742.601462536909	47.224649070323\\
742.097148745079	47.0842762953286\\
741.717488927165	46.9753408846414\\
741.392924151083	46.8890638264271\\
741.115734190453	46.8191015495357\\
740.905557947835	46.7614261293075\\
740.73067636565	46.713554692953\\
740.557301919291	46.672590813522\\
740.411125123032	46.6374743797146\\
740.285602239173	46.607089650319\\
740.178157295768	46.58048977857\\
740.072327140748	46.5567403734722\\
739.982129675197	46.5356275059848\\
739.90105960876	46.5166758094132\\
739.814945250984	46.4996366356887\\
739.768470103346	46.4841668767399\\
739.706662155512	46.4700151030807\\
739.644731176181	46.4570267282312\\
739.583300012303	46.4450908807746\\
739.541646161417	46.434065177231\\
739.494717335138	46.4237878339966\\
739.463459645669	46.4142771515298\\
739.419106791339	46.4053788615115\\
739.393077940453	46.3971327759323\\
739.354692113681	46.3893221070979\\
739.322134904035	46.3820170551987\\
739.281934362697	46.3751640672308\\
739.273345226378	46.3686360703577\\
739.232875553642	46.3625302739442\\
739.212298535925	46.356672650148\\
739.193559301181	46.3511844263065\\
739.174781619094	46.3459357611876\\
739.154942790354	46.3409633742724\\
739.128337229331	46.3362087730061\\
739.108667568898	46.3316365591762\\
739.091235543799	46.3273175844138\\
739.072373277559	46.3231817972047\\
739.052496001476	46.3192686968839\\
739.039047121063	46.3154829495685\\
739.015402005413	46.3118522594847\\
739.00622078002	46.3084139969839\\
738.989280880906	46.3051057066518\\
738.99151851624	46.3019285898965\\
738.969718873032	46.2988464168669\\
738.961345041831	46.2958686631784\\
738.945381705217	46.2930323938268\\
738.943659264272	46.2902729079039\\
738.932178887795	46.2876072301415\\
738.929972010335	46.2850790456343\\
738.91488527313	46.2825961151476\\
738.901390255906	46.2801904504614\\
738.89303949311	46.2778512128454\\
738.885334645669	46.2756143661877\\
738.879605991634	46.273455407928\\
738.866533895177	46.2713388851541\\
738.856491449311	46.2692998818533\\
738.85387703002	46.2673371828206\\
738.850316806102	46.2654428425372\\
738.845134104331	46.2636132279154\\
738.829755167323	46.2618313846394\\
738.829239972933	46.2601055000128\\
738.826671690453	46.2584344185246\\
738.817836491142	46.2568095432472\\
738.816260150098	46.2552237843928\\
738.796890378937	46.2536583150115\\
738.789654589075	46.2521473005748\\
738.79486804872	46.2506788536838\\
738.785448449803	46.2492549777296\\
738.773883489173	46.2478848938377\\
738.775782787894	46.2465310525459\\
738.770546259842	46.2452246012649\\
738.774352546752	46.2439627346518\\
738.762272391732	46.2427063075115\\
738.766724593996	46.2415057916123\\
738.756443774606	46.2403078319364\\
738.738781065453	46.239147536558\\
738.75457523376	46.2380166218731\\
738.74727792815	46.2369034312124\\
738.744109867126	46.2358214326792\\
738.752037709154	46.2347547956571\\
738.738496555118	46.2337223664385\\
738.741203248032	46.2327268581385\\
738.734044352854	46.2317344333739\\
738.735167015256	46.2307752814141\\
738.719888041339	46.2298277526084\\
738.728584830217	46.2288996063547\\
738.718165600394	46.2279723991391\\
738.715274360236	46.2270866297956\\
738.710929810532	46.2262406442606\\
738.70949956939	46.2254141458388\\
738.707946296752	46.2246150390921\\
738.696558193898	46.2238121851852\\
738.707177349902	46.2230239573945\\
738.69562007874	46.2222530884939\\
738.694312869094	46.2214970659514\\
};\label{gaussian};

\addplot [color=mycolor4,solid,mark=triangle,mark options={solid},mark repeat = 11,mark phase = 1]
  table[row sep=crcr]{%
758.039892962598	48.9622582667373\\
757.930271899606	48.9591725395369\\
757.678426427165	48.9535147488427\\
757.318136380413	48.9447233527593\\
756.82458015502	48.9325572139527\\
756.277382197342	48.9170647118\\
755.676234928642	48.8985156665038\\
755.049843134842	48.8773415507238\\
754.420344795768	48.8541438960255\\
753.805756336122	48.8295181408916\\
753.230061208169	48.8040830505996\\
752.700372170276	48.778329233893\\
752.209876353346	48.7526429970405\\
751.760549950787	48.7272325444918\\
751.350885826772	48.7022693706385\\
750.97543214813	48.6778173057584\\
750.628867802658	48.6538434387089\\
750.307694082185	48.630388290134\\
749.989250123032	48.6074345437234\\
749.722640871063	48.5849127288762\\
749.471910371555	48.5628553134215\\
749.219926488681	48.5412101341976\\
748.993725393701	48.519995259168\\
748.78890871063	48.4991499902762\\
748.582738681102	48.4787613586453\\
748.39108636811	48.4586826937275\\
748.208653727854	48.4390556963735\\
748.036540354331	48.4197342154827\\
747.884381151575	48.4007043026333\\
747.736205093504	48.3819651212256\\
747.585399237205	48.3634647062624\\
747.428672490158	48.3450707006197\\
747.306094672736	48.3273484087257\\
747.174274114173	48.3097697267625\\
747.068590059055	48.2926376046722\\
746.940106729823	48.2754012632529\\
746.822234867126	48.2588294980585\\
746.727093073327	48.2425549992658\\
746.630182701772	48.2264338363768\\
746.520154096949	48.2108084592596\\
746.427703617126	48.1955621833331\\
746.349055733268	48.1801631841379\\
746.258220041831	48.1651849669828\\
746.161663385827	48.1505889084126\\
746.080762487697	48.1359048322925\\
746.009350393701	48.1216575734686\\
745.913716473917	48.1079153807537\\
745.845557025098	48.0943349774943\\
745.779120017224	48.0808677507204\\
745.712336983268	48.0674290734519\\
745.651736281988	48.0545079951789\\
745.573211429626	48.0418021470252\\
745.527782049705	48.0294039441157\\
745.453809362697	48.0168794929735\\
745.409887118602	48.0051145655935\\
745.341197096457	47.9934757900187\\
745.282987819882	47.9817461795066\\
745.230499507874	47.9702997525873\\
745.174397145669	47.9589926892432\\
745.136826402559	47.948045468653\\
745.074810839075	47.9375488397091\\
745.027328371063	47.9268100727009\\
744.983267716535	47.9163469366291\\
744.929333784449	47.9056837209483\\
744.890048289862	47.8955151556965\\
744.849363312008	47.8853698839941\\
744.80880136565	47.8753478765807\\
744.768570066437	47.8655731568319\\
744.715635765256	47.8559814034295\\
744.683839812992	47.8465309161801\\
744.642178272638	47.8373619553686\\
744.589720718504	47.8282445017712\\
744.563645730807	47.8190866649179\\
744.525836614173	47.8102303779375\\
744.485912893701	47.8013667547661\\
744.443082554134	47.792703257886\\
744.419260580709	47.7844665481271\\
744.373777374508	47.7763955493188\\
744.338913324311	47.7687047551651\\
744.323634350394	47.7606076208342\\
744.28597902313	47.7528216898693\\
744.262318528543	47.7454134028361\\
744.223109928642	47.7378416509591\\
744.176065760335	47.7305240162329\\
744.164039431594	47.7232676268393\\
744.129544475886	47.7162502988362\\
744.097917691929	47.7093536911379\\
744.063738004429	47.7025914198467\\
744.052173043799	47.6959870807249\\
744.02795890748	47.6893843685647\\
744.000084584154	47.682934349791\\
743.985674520177	47.676813470589\\
743.953463336614	47.6706573202902\\
743.933255413386	47.6643722272126\\
743.899306409941	47.6580075260059\\
743.865788016732	47.6520678846818\\
743.867202878937	47.6461155679983\\
743.818651574803	47.6405157899056\\
743.801350270669	47.6349977693954\\
743.804556779035	47.6293100168082\\
};
\label{BM3D};

\addplot [color=mycolor3,solid,mark=o,mark options={solid},mark repeat = 8,mark phase = 1]
table[row sep=crcr]{%
	758.608121616634	48.9532378175578\\
	757.077348363681	48.9093199661386\\
	755.29363773376	48.9029525243455\\
	754.055479515256	48.8962657355409\\
	752.999592458169	48.8820289247319\\
	751.999876968504	48.8581593397891\\
	751.087859867126	48.8249044978022\\
	750.232106606791	48.7840598037255\\
	749.511687992126	48.7381786117334\\
	748.879790538878	48.6897159669122\\
	748.378421813484	48.6405957120973\\
	747.940368171752	48.5922454836879\\
	747.569912647638	48.5455219796204\\
	747.293437807579	48.5009965833196\\
	747.072488619587	48.4588942125255\\
	746.880221149114	48.419284453064\\
	746.727515994094	48.3821967320403\\
	746.602100762795	48.3474818055737\\
	746.507082000492	48.3150739050799\\
	746.414508489173	48.2847966763349\\
	746.341273991142	48.2564845529353\\
	746.298874261811	48.2300160364534\\
	746.244025282972	48.2052532307123\\
	746.211667999508	48.1820525886808\\
	746.156411478839	48.1602864643329\\
	746.12839105561	48.1398607858454\\
	746.115249753937	48.1206523333027\\
	746.096548966535	48.1025777713009\\
	746.07807117372	48.0855482246066\\
	746.091942974902	48.0694937569302\\
	746.063222810039	48.0543394623761\\
	746.060639148622	48.0400063576434\\
	746.041707677165	48.0264414213522\\
	746.049435593012	48.0135906772271\\
	746.040415846457	48.0014242174356\\
	746.041500061516	47.9898693200546\\
	746.03890871063	47.9788876473659\\
	746.03885488435	47.9684626944583\\
	746.051465612697	47.9585251703339\\
	746.039662278543	47.949066940882\\
	746.051288754921	47.9400555832049\\
	746.050858144685	47.9314626304853\\
	746.045721579724	47.923253198525\\
	746.03967765748	47.9154141958566\\
	746.045667753445	47.9079229887802\\
	746.050696665846	47.9007467145823\\
	746.050081508366	47.8938930583257\\
	746.038885642224	47.8873178202334\\
	746.042153666339	47.8810219321338\\
	746.054541400098	47.8749749968602\\
	746.054787463091	47.8691805120875\\
	746.057932455709	47.8636129970284\\
	746.058693713091	47.8582640487144\\
	746.04821296752	47.8531173506565\\
	746.057517224409	47.8481797968946\\
	746.054733636811	47.8434215401195\\
	746.062461552658	47.8388483647488\\
	746.064653051181	47.8344295759521\\
	746.06342273622	47.8301816611215\\
	746.05602546752	47.8260903114347\\
	746.061623400591	47.8221414431309\\
	746.060262364665	47.8183279258303\\
	746.05807855561	47.8146500802962\\
	746.067467396654	47.8111013911743\\
	746.073611281988	47.8076607996106\\
	746.062323142224	47.8043418722523\\
	746.069151390256	47.8011365425887\\
	746.087467704232	47.7980345091629\\
	746.075895054134	47.7950263509301\\
	746.073288324311	47.7921176464685\\
	746.084484190453	47.78930227101\\
	746.081915907972	47.7865755996842\\
	746.081316129429	47.7839339324316\\
	746.083338459646	47.7813747619446\\
	746.09539554626	47.7788951887795\\
	746.094611220472	47.7764843253823\\
	746.090335875984	47.7741528269761\\
	746.09662586122	47.7718892711435\\
	746.101147268701	47.7696885799745\\
	746.1044921875	47.767557731311\\
	746.098801980807	47.7654777509984\\
	746.094818836122	47.7634619913628\\
	746.103115772638	47.7614911178915\\
	746.102754367618	47.7595902005356\\
	746.093934547244	47.7577375195012\\
	746.09914800689	47.7559317999021\\
	746.109874815453	47.7541829024856\\
	746.103338767224	47.7524754135582\\
	746.09966320128	47.7508138287807\\
	746.101070374016	47.7492032444819\\
	746.092289000984	47.7476325298098\\
	746.09827140748	47.7460997693413\\
	746.099832369587	47.7446079926047\\
	746.100101500984	47.7431509097951\\
	746.110851377953	47.7417324588328\\
	746.106014702264	47.7403532880478\\
	746.110120878445	47.7390056943637\\
	746.106722133366	47.7376891350925\\
	746.10413078248	47.7364128840252\\
	746.108929010827	47.7351601704802\\
};
\label{NLM};

\addplot [color=mycolor5,solid,mark=diamond,mark options={solid},mark repeat = 11,mark phase = 1]
  table[row sep=crcr]{%
757.452809731791	48.9609619563465\\
757.040431225394	48.9590130404662\\
756.608013964075	48.9567954997526\\
756.181955893209	48.9543754522261\\
755.796182947835	48.9518089980595\\
755.425273745079	48.9491344745402\\
755.077463705709	48.9463796350714\\
754.765763410433	48.9435579429659\\
754.449457123524	48.9406843438755\\
754.164531557579	48.9377612851017\\
753.903704785925	48.9347989529133\\
753.633650652067	48.9318170568784\\
753.388856422244	48.9288107982518\\
753.145523191437	48.9257805303276\\
752.930541031004	48.9227364842105\\
752.710968257874	48.91968807462\\
752.497385580709	48.9166310675201\\
752.293706938976	48.91357301105\\
752.110889825295	48.9105110947056\\
751.928833968996	48.9074510629937\\
751.74126476378	48.9043922033514\\
751.553718626968	48.9013394083711\\
751.395084891732	48.8982884851051\\
751.247485543799	48.8952482686593\\
751.07874015748	48.8922143292989\\
750.937984436516	48.8891918992498\\
750.789539247047	48.8861727580734\\
750.652674397146	48.8831702540097\\
750.508819820374	48.8801801508348\\
750.390994094488	48.877201111506\\
750.26629398376	48.8742378864196\\
750.137664554626	48.8712776615899\\
750.022799274114	48.8683355699778\\
749.907926304134	48.8654069337495\\
749.792722687008	48.8625013112343\\
749.688084399606	48.8596088729023\\
749.578378752461	48.8567313963564\\
749.483782910925	48.8538683716606\\
749.385173166831	48.8510221826402\\
749.279289185532	48.8481866896026\\
749.193682332677	48.8453689843439\\
749.100393700787	48.8425692270715\\
748.999446358268	48.8397908000169\\
748.911686454232	48.8370237265994\\
748.81424550935	48.8342734653233\\
748.735628383366	48.8315353691344\\
748.650705893209	48.8288164465326\\
748.562461552658	48.8261147214324\\
748.495909202756	48.8234293053136\\
748.40655757874	48.8207614215767\\
748.338228961614	48.8181080595439\\
748.259012057087	48.8154738429164\\
748.196765809547	48.8128528708983\\
748.113035187008	48.8102432254939\\
748.050665907972	48.8076539004093\\
747.988873339075	48.8050729571917\\
747.919199064961	48.8025175013026\\
747.84909418061	48.7999786517258\\
747.783756766732	48.797451730676\\
747.735636072835	48.7949386233152\\
747.674827755906	48.7924446315065\\
747.611474224902	48.7899657645643\\
747.551011934055	48.7875003465068\\
747.49614757628	48.7850500910099\\
747.436431163878	48.7826196052965\\
747.379336860236	48.7801974493377\\
747.337990588091	48.7777985340003\\
747.288785679134	48.7754107492139\\
747.23483636811	48.7730406991964\\
747.182471087598	48.7706811964993\\
747.111458845965	48.7683340609515\\
747.067621186024	48.7660024019788\\
747.025675135335	48.7636915690568\\
746.982998585138	48.7613946193709\\
746.93673105315	48.7591087418854\\
746.897245632382	48.7568385991263\\
746.846656619094	48.754582257607\\
746.80268054872	48.7523319188227\\
746.752729761319	48.7500987405775\\
746.717212106299	48.7478793261572\\
746.67850332185	48.7456740228767\\
746.637556902067	48.7434782900877\\
746.577440637303	48.7413019783317\\
746.542645792323	48.7391371118291\\
746.503168061024	48.7369862440501\\
746.480476439468	48.7348460389044\\
746.438468873032	48.7327203581381\\
746.389979084646	48.7306028054277\\
746.365449680118	48.7284968949797\\
746.321850393701	48.7264120270105\\
746.282203494094	48.7243338415121\\
746.248746616634	48.7222639024723\\
746.211614173228	48.720205436835\\
746.186123585138	48.7181654076115\\
746.14864511565	48.7161390980044\\
746.111366572342	48.7141194569416\\
746.085191621555	48.7121153461847\\
746.050058439961	48.7101215655485\\
746.034187376968	48.7081379917669\\
745.997393270177	48.7061657178345\\
};
\label{GIF};

\addplot [color=mycolor6,solid,mark=+,mark options={solid,scale=3,thick},only marks]
  table[row sep=crcr]{%
738.526859313484	46.166222065342\\
};
\label{trunc};

\addplot [color=mycolor2,solid,mark=square,mark options={solid},mark repeat = 11,mark phase = 1]
table[row sep=crcr]{%
	757.381020853839	48.9605201492037\\
	756.974263348917	48.9576137038724\\
	756.646599717028	48.954326482023\\
	756.364350086122	48.950817100077\\
	756.125507504921	48.9471803096613\\
	755.936546505906	48.9434820282419\\
	755.761634165846	48.9397609524437\\
	755.605168860728	48.9360572349982\\
	755.463967150591	48.932400117227\\
	755.335968257874	48.9287897840792\\
	755.226739357776	48.9252431515114\\
	755.122485543799	48.9217607015256\\
	755.035594549705	48.9183641340663\\
	754.945758489173	48.9150499318341\\
	754.8642578125	48.9118176252294\\
	754.790800319882	48.9086706770303\\
	754.710622231791	48.9056071296521\\
	754.635288508858	48.902622368232\\
	754.579747477854	48.899724495954\\
	754.528820127953	48.8969044530236\\
	754.456039308563	48.8941628859617\\
	754.406280757874	48.8914982236044\\
	754.358959768701	48.8888995994714\\
	754.300696665846	48.8863789253638\\
	754.261111281988	48.8839296931895\\
	754.208384596457	48.8815507759454\\
	754.174697034941	48.8792364078257\\
	754.121878075787	48.8769860086634\\
	754.085583784449	48.8747944961524\\
	754.035440760335	48.8726713261618\\
	754.016094057579	48.8706061175544\\
	753.977992741142	48.8685959730786\\
	753.947603961614	48.8666404964251\\
	753.909625676673	48.8647336269674\\
	753.884081262303	48.8628807098521\\
	753.862842950295	48.8610754751014\\
	753.827340674213	48.859316387031\\
	753.792868786909	48.8576074410355\\
	753.76624015748	48.8559387994607\\
	753.734959399606	48.8543168702192\\
	753.719096026083	48.8527395203593\\
	753.686861774114	48.8511948248934\\
	753.664500799705	48.8496867545213\\
	753.632558747539	48.8482204744603\\
	753.602992741142	48.846790729734\\
	753.587598425197	48.8453979542388\\
	753.560177780512	48.8440407690893\\
	753.542476624016	48.8427169104417\\
	753.52307609498	48.8414206290582\\
	753.497270238681	48.8401592721599\\
	753.482721764272	48.8389261936048\\
	753.456731360728	48.8377214185447\\
	753.430110420768	48.8365469996989\\
	753.426273375984	48.8353955047825\\
	753.397084153543	48.834275340812\\
	753.382304995079	48.8331797594111\\
	753.37186269685	48.8321081375046\\
	753.34884042815	48.8310626889531\\
	753.337213951772	48.8300392644743\\
	753.326248769685	48.8290362765648\\
	753.303787832185	48.8280606496034\\
	753.287155511811	48.8271024003144\\
	753.280950110728	48.8261623092974\\
	753.258450725886	48.8252495005356\\
	753.243194820374	48.8243538894269\\
	753.243586983268	48.8234764091836\\
	753.214751476378	48.8226115153165\\
	753.206231545276	48.8217699006962\\
	753.198426734744	48.8209484401498\\
	753.178949311024	48.8201399497133\\
	753.162578432579	48.8193490469624\\
	753.166653850886	48.8185774107291\\
	753.146968811516	48.8178222954483\\
	753.140025221457	48.8170799272293\\
	753.13003660187	48.8163543716193\\
	753.123346764272	48.8156415589779\\
	753.097871555118	48.814949941149\\
	753.096702755906	48.8142680837737\\
	753.086368110236	48.8135947146067\\
	753.074787770669	48.8129352817844\\
	753.07092765748	48.8122901752496\\
	753.05627921998	48.8116532111878\\
	753.043445497047	48.8110292733934\\
	753.037109375	48.810419862444\\
	753.01608636811	48.8098220402627\\
	753.01593257874	48.8092360733069\\
	753.010811392717	48.8086604798982\\
	752.992156742126	48.8080984693983\\
	752.983106237697	48.8075445641979\\
	752.974901574803	48.8070010343566\\
	752.965528112697	48.806467224902\\
	752.966758427658	48.8059409933117\\
	752.952886626476	48.8054244590068\\
	752.940421998032	48.8049177120099\\
	752.927026943898	48.8044219659854\\
	752.925181471457	48.8039319252211\\
	752.92022945374	48.8034516174242\\
	752.912140132874	48.802980730223\\
	752.898929625984	48.8025181718264\\
	752.899644746555	48.8020607535835\\
};
\label{AWF};

\addplot [color=mycolor1,solid,mark=asterisk,mark options={solid,scale=3,thick},only marks]
table[row sep=crcr]{%
	758.434685654528	48.9641984944098\\
};
\label{none};

\nextgroupplot[%
width=\figurewidth ,
height=0.5\figurewidth ,
scale only axis,
xmin=737,
xmax=760,
xlabel={File size LP [kB]},
xmajorgrids,
ymin=0.9892,
ymax=0.995,
title={$\xi:$ $100\longleftarrow 1$},
title style={at={(1,1.1)},anchor=north east},
y tick label style={
	/pgf/number format/.cd,
	fixed,
	fixed zerofill,
	precision=4,
	/tikz/.cd
},
ylabel={$\text{SSIM}_{\text{LP}_t}$},
ymajorgrids,
axis background/.style={fill=white}
]

\addplot [color=mycolor7,dashed,mark=Mercedes star flipped,mark options={solid},mark repeat = 11,mark phase = 1]
table[row sep=crcr]{%
752.783372293307	0.99431721125251\\
748.09842519685	0.9936631245338\\
745.649514025591	0.992890551613193\\
744.201417937992	0.992216749654831\\
743.273560531496	0.991683997009985\\
742.601462536909	0.991275440237488\\
742.097148745079	0.990965544090037\\
741.717488927165	0.990729075939416\\
741.392924151083	0.990546638135394\\
741.115734190453	0.990403941341668\\
740.905557947835	0.99029123406986\\
740.73067636565	0.990202402173487\\
740.557301919291	0.990129911477535\\
740.411125123032	0.990070836825791\\
740.285602239173	0.990022611481637\\
740.178157295768	0.989982478235978\\
740.072327140748	0.989948655819432\\
739.982129675197	0.989920292834424\\
739.90105960876	0.989896027458026\\
739.814945250984	0.989875385812511\\
739.768470103346	0.989857591927749\\
739.706662155512	0.989842158173755\\
739.644731176181	0.989828703405909\\
739.583300012303	0.989816974383811\\
739.541646161417	0.9898066101571\\
739.494717335138	0.989797403074393\\
739.463459645669	0.989789248591788\\
739.419106791339	0.989781940324903\\
739.393077940453	0.989775485776497\\
739.354692113681	0.989769624136705\\
739.322134904035	0.989764373251143\\
739.281934362697	0.989759582939486\\
739.273345226378	0.989755228906686\\
739.232875553642	0.989751332785093\\
739.212298535925	0.989747655093022\\
739.193559301181	0.989744321423604\\
739.174781619094	0.989741244252415\\
739.154942790354	0.989738494110652\\
739.128337229331	0.989735871326872\\
739.108667568898	0.98973344923867\\
739.091235543799	0.989731186390989\\
739.072373277559	0.989729091511409\\
739.052496001476	0.989727174992771\\
739.039047121063	0.989725342291731\\
739.015402005413	0.989723659595045\\
739.00622078002	0.989722111474879\\
738.989280880906	0.989720639175597\\
738.99151851624	0.989719272489159\\
738.969718873032	0.989717918335726\\
738.961345041831	0.989716613880198\\
738.945381705217	0.989715421954967\\
738.943659264272	0.989714273038612\\
738.932178887795	0.989713158224094\\
738.929972010335	0.989712137688875\\
738.91488527313	0.989711142884512\\
738.901390255906	0.989710194901405\\
738.89303949311	0.989709263993619\\
738.885334645669	0.989708396957037\\
738.879605991634	0.989707562812696\\
738.866533895177	0.989706715945297\\
738.856491449311	0.989705957670835\\
738.85387703002	0.989705270020729\\
738.850316806102	0.989704589072546\\
738.845134104331	0.989703950129027\\
738.829755167323	0.989703330638477\\
738.829239972933	0.989702777469022\\
738.826671690453	0.989702187986901\\
738.817836491142	0.989701636935139\\
738.816260150098	0.989701097980093\\
738.796890378937	0.989700550652225\\
738.789654589075	0.989699986634926\\
738.79486804872	0.989699467182427\\
738.785448449803	0.989698983872532\\
738.773883489173	0.9896985540216\\
738.775782787894	0.989698103313499\\
738.770546259842	0.989697692156052\\
738.774352546752	0.989697321867101\\
738.762272391732	0.989696914810784\\
738.766724593996	0.989696552962529\\
738.756443774606	0.989696160416794\\
738.738781065453	0.989695788168433\\
738.75457523376	0.989695419679809\\
738.74727792815	0.989695060896702\\
738.744109867126	0.989694740766576\\
738.752037709154	0.989694431959088\\
738.738496555118	0.989694148157428\\
738.741203248032	0.989693856278588\\
738.734044352854	0.989693539506162\\
738.735167015256	0.989693236993608\\
738.719888041339	0.989692963475621\\
738.728584830217	0.989692677391365\\
738.718165600394	0.989692367240007\\
738.715274360236	0.989692085719108\\
738.710929810532	0.989691823307195\\
738.70949956939	0.989691563295074\\
738.707946296752	0.989691321213526\\
738.696558193898	0.989691055925758\\
738.707177349902	0.989690811999546\\
738.69562007874	0.989690568178456\\
738.694312869094	0.989690329294126\\
};

\addplot [color=mycolor4,solid,mark=triangle,mark options={solid},mark repeat = 11,mark phase = 1]
table[row sep=crcr]{%
	758.039892962598	0.994470658049905\\
	757.930271899606	0.994469781777303\\
	757.678426427165	0.994466265331743\\
	757.318136380413	0.994458511667376\\
	756.82458015502	0.994445337863166\\
	756.277382197342	0.994426147187198\\
	755.676234928642	0.994400925250864\\
	755.049843134842	0.994370091932459\\
	754.420344795768	0.994334617799309\\
	753.805756336122	0.99429547129473\\
	753.230061208169	0.994253854834174\\
	752.700372170276	0.99421076795795\\
	752.209876353346	0.994166977973947\\
	751.760549950787	0.994123062312484\\
	751.350885826772	0.994079351120466\\
	750.97543214813	0.994036049582044\\
	750.628867802658	0.993993153500529\\
	750.307694082185	0.993950825120064\\
	749.989250123032	0.993909002895976\\
	749.722640871063	0.993867633586209\\
	749.471910371555	0.993826730419758\\
	749.219926488681	0.993786355515675\\
	748.993725393701	0.993746484077626\\
	748.78890871063	0.993707065096276\\
	748.582738681102	0.993668271528037\\
	748.39108636811	0.993629828403088\\
	748.208653727854	0.993592034718272\\
	748.036540354331	0.993554642433068\\
	747.884381151575	0.993517665388857\\
	747.736205093504	0.993481092809662\\
	747.585399237205	0.993444968958063\\
	747.428672490158	0.99340908530417\\
	747.306094672736	0.993374067216465\\
	747.174274114173	0.99333929203176\\
	747.068590059055	0.99330523177549\\
	746.940106729823	0.993271154469486\\
	746.822234867126	0.993238111898459\\
	746.727093073327	0.993205486023503\\
	746.630182701772	0.993173314417071\\
	746.520154096949	0.99314177753832\\
	746.427703617126	0.993110927590227\\
	746.349055733268	0.99307999823829\\
	746.258220041831	0.993049805384693\\
	746.161663385827	0.993020222846727\\
	746.080762487697	0.992990617037869\\
	746.009350393701	0.992961775620008\\
	745.913716473917	0.992933702211354\\
	745.845557025098	0.992906111817028\\
	745.779120017224	0.992878787710175\\
	745.712336983268	0.992851566552704\\
	745.651736281988	0.992825261788351\\
	745.573211429626	0.992799334061887\\
	745.527782049705	0.992774014425875\\
	745.453809362697	0.992748577995419\\
	745.409887118602	0.992724347636329\\
	745.341197096457	0.99270034092723\\
	745.282987819882	0.992676399547418\\
	745.230499507874	0.992652927634798\\
	745.174397145669	0.992629743033882\\
	745.136826402559	0.992607262349927\\
	745.074810839075	0.992585525121902\\
	745.027328371063	0.992563537023983\\
	744.983267716535	0.992542145087391\\
	744.929333784449	0.992520320858981\\
	744.890048289862	0.992499561434469\\
	744.849363312008	0.992478519611089\\
	744.80880136565	0.992457945924352\\
	744.768570066437	0.992437940703577\\
	744.715635765256	0.99241804392645\\
	744.683839812992	0.992398646931918\\
	744.642178272638	0.992379770657205\\
	744.589720718504	0.992360794370988\\
	744.563645730807	0.992342079632193\\
	744.525836614173	0.992323922565086\\
	744.485912893701	0.992305817225614\\
	744.443082554134	0.99228794190998\\
	744.419260580709	0.992270772715099\\
	744.373777374508	0.992253943299778\\
	744.338913324311	0.99223794389618\\
	744.323634350394	0.992221063096848\\
	744.28597902313	0.992204798547212\\
	744.262318528543	0.992189333221\\
	744.223109928642	0.992173592557281\\
	744.176065760335	0.992158293609951\\
	744.164039431594	0.992143190386952\\
	744.129544475886	0.992128604198835\\
	744.097917691929	0.992114319258383\\
	744.063738004429	0.992100283444141\\
	744.052173043799	0.992086582043788\\
	744.02795890748	0.992072913881477\\
	744.000084584154	0.992059614382063\\
	743.985674520177	0.992046936924383\\
	743.953463336614	0.992034125233877\\
	743.933255413386	0.992021119408818\\
	743.899306409941	0.992007996902635\\
	743.865788016732	0.991995729104459\\
	743.867202878937	0.991983430920426\\
	743.818651574803	0.991971821019933\\
	743.801350270669	0.991960367259965\\
	743.804556779035	0.991948689498966\\
};

\addplot [color=mycolor3,solid,mark=o,mark options={solid},mark repeat = 8,mark phase = 1]
table[row sep=crcr]{%
	758.608121616634	0.994454384880541\\
	757.077348363681	0.994389639804636\\
	755.29363773376	0.994386369669719\\
	754.055479515256	0.994377470340458\\
	752.999592458169	0.994354149267087\\
	751.999876968504	0.994313316220762\\
	751.087859867126	0.994255083083025\\
	750.232106606791	0.994182212723079\\
	749.511687992126	0.994099016263306\\
	748.879790538878	0.994010026386508\\
	748.378421813484	0.993918907950923\\
	747.940368171752	0.99382856237282\\
	747.569912647638	0.993740855694232\\
	747.293437807579	0.993656989664955\\
	747.072488619587	0.99357755410385\\
	746.880221149114	0.993502865714746\\
	746.727515994094	0.993432973883929\\
	746.602100762795	0.99336763185942\\
	746.507082000492	0.993306774031049\\
	746.414508489173	0.993250085123622\\
	746.341273991142	0.993197216440045\\
	746.298874261811	0.99314795514265\\
	746.244025282972	0.993102013558891\\
	746.211667999508	0.993059139674084\\
	746.156411478839	0.993019070452654\\
	746.12839105561	0.992981563366907\\
	746.115249753937	0.992946409279803\\
	746.096548966535	0.992913469685817\\
	746.07807117372	0.992882533815461\\
	746.091942974902	0.992853455716161\\
	746.063222810039	0.992826089691739\\
	746.060639148622	0.992800296025297\\
	746.041707677165	0.992775937489527\\
	746.049435593012	0.992752941869262\\
	746.040415846457	0.992731220848875\\
	746.041500061516	0.992710661996224\\
	746.03890871063	0.992691184787564\\
	746.03885488435	0.992672719189441\\
	746.051465612697	0.992655157391412\\
	746.039662278543	0.992638493470984\\
	746.051288754921	0.992622638949799\\
	746.050858144685	0.992607564025571\\
	746.045721579724	0.992593197304705\\
	746.03967765748	0.992579489098285\\
	746.045667753445	0.992566414647848\\
	746.050696665846	0.9925539186796\\
	746.050081508366	0.992542009855161\\
	746.038885642224	0.992530617870053\\
	746.042153666339	0.992519705488028\\
	746.054541400098	0.992509268881498\\
	746.054787463091	0.992499272674374\\
	746.057932455709	0.992489676391968\\
	746.058693713091	0.992480476968574\\
	746.04821296752	0.992471638576438\\
	746.057517224409	0.992463167984908\\
	746.054733636811	0.992455012307073\\
	746.062461552658	0.992447193092066\\
	746.064653051181	0.992439640726175\\
	746.06342273622	0.992432395335392\\
	746.05602546752	0.992425419255256\\
	746.061623400591	0.992418689027021\\
	746.060262364665	0.992412213410229\\
	746.05807855561	0.992405978079719\\
	746.067467396654	0.992399948791425\\
	746.073611281988	0.992394119488609\\
	746.062323142224	0.992388507853455\\
	746.069151390256	0.992383084258817\\
	746.087467704232	0.992377840156948\\
	746.075895054134	0.99237277437091\\
	746.073288324311	0.992367865275392\\
	746.084484190453	0.992363122261145\\
	746.081915907972	0.992358538706656\\
	746.081316129429	0.992354102010939\\
	746.083338459646	0.992349802187835\\
	746.09539554626	0.992345643935441\\
	746.094611220472	0.99234160529249\\
	746.090335875984	0.99233770123514\\
	746.09662586122	0.992333918858534\\
	746.101147268701	0.992330240554315\\
	746.1044921875	0.99232667386964\\
	746.098801980807	0.992323197988368\\
	746.094818836122	0.992319839323522\\
	746.103115772638	0.992316560105331\\
	746.102754367618	0.992313387628051\\
	746.093934547244	0.992310301469052\\
	746.09914800689	0.992307300710812\\
	746.109874815453	0.992304388312196\\
	746.103338767224	0.992301553899624\\
	746.09966320128	0.992298799528231\\
	746.101070374016	0.992296132514275\\
	746.092289000984	0.992293533972549\\
	746.09827140748	0.992291005946491\\
	746.099832369587	0.992288539749466\\
	746.100101500984	0.992286129699547\\
	746.110851377953	0.992283784713217\\
	746.106014702264	0.992281514531909\\
	746.110120878445	0.992279293285901\\
	746.106722133366	0.992277118347903\\
	746.10413078248	0.99227501810335\\
	746.108929010827	0.992272952076212\\
};

\addplot [color=mycolor5,solid,mark=diamond,mark options={solid},mark repeat = 11,mark phase = 1]
table[row sep=crcr]{%
	757.452809731791	0.994465502178838\\
	757.040431225394	0.994461346933884\\
	756.608013964075	0.994456591829416\\
	756.181955893209	0.99445144897495\\
	755.796182947835	0.994446055224043\\
	755.425273745079	0.994440498736784\\
	755.077463705709	0.994434799752409\\
	754.765763410433	0.994429002369292\\
	754.449457123524	0.994423140523498\\
	754.164531557579	0.994417200640622\\
	753.903704785925	0.994411223287948\\
	753.633650652067	0.994405218190131\\
	753.388856422244	0.994399198047662\\
	753.145523191437	0.994393160220441\\
	752.930541031004	0.994387120092725\\
	752.710968257874	0.994381080093054\\
	752.497385580709	0.994375027243622\\
	752.293706938976	0.994368996987687\\
	752.110889825295	0.994362963624629\\
	751.928833968996	0.994356963235625\\
	751.74126476378	0.994350971641375\\
	751.553718626968	0.994345015057749\\
	751.395084891732	0.994339066074802\\
	751.247485543799	0.994333159742972\\
	751.07874015748	0.994327277069764\\
	750.937984436516	0.99432142444789\\
	750.789539247047	0.994315582252475\\
	750.652674397146	0.994309795484557\\
	750.508819820374	0.994304031827189\\
	750.390994094488	0.994298302513091\\
	750.26629398376	0.994292605824921\\
	750.137664554626	0.994286929168541\\
	750.022799274114	0.994281287600869\\
	749.907926304134	0.994275677493418\\
	749.792722687008	0.994270132605771\\
	749.688084399606	0.994264607741378\\
	749.578378752461	0.99425911833478\\
	749.483782910925	0.994253665235314\\
	749.385173166831	0.994248255304441\\
	749.279289185532	0.994242875213711\\
	749.193682332677	0.994237532570271\\
	749.100393700787	0.994232225477125\\
	748.999446358268	0.994226962962076\\
	748.911686454232	0.994221728922168\\
	748.81424550935	0.994216531999133\\
	748.735628383366	0.994211365240321\\
	748.650705893209	0.994206235127219\\
	748.562461552658	0.994201148552508\\
	748.495909202756	0.99419608679263\\
	748.40655757874	0.99419107228778\\
	748.338228961614	0.994186084106695\\
	748.259012057087	0.994181141387976\\
	748.196765809547	0.994176223546923\\
	748.113035187008	0.994171330688518\\
	748.050665907972	0.994166485619614\\
	747.988873339075	0.994161658471227\\
	747.919199064961	0.994156873058469\\
	747.84909418061	0.994152128369529\\
	747.783756766732	0.994147394909292\\
	747.735636072835	0.994142701948882\\
	747.674827755906	0.994138054140155\\
	747.611474224902	0.994133428558154\\
	747.551011934055	0.994128828210197\\
	747.49614757628	0.994124265629332\\
	747.436431163878	0.994119738906229\\
	747.379336860236	0.994115240589274\\
	747.337990588091	0.994110780861763\\
	747.288785679134	0.994106338751916\\
	747.23483636811	0.994101934971194\\
	747.182471087598	0.994097552272106\\
	747.111458845965	0.994093201785115\\
	747.067621186024	0.994088880627359\\
	747.025675135335	0.994084604044769\\
	746.982998585138	0.994080355379649\\
	746.93673105315	0.994076126801092\\
	746.897245632382	0.994071926393705\\
	746.846656619094	0.994067750119159\\
	746.80268054872	0.99406358832081\\
	746.752729761319	0.994059459359829\\
	746.717212106299	0.994055365479658\\
	746.67850332185	0.994051292753131\\
	746.637556902067	0.994047245650776\\
	746.577440637303	0.994043227879138\\
	746.542645792323	0.994039234281253\\
	746.503168061024	0.994035267815776\\
	746.480476439468	0.994031324660088\\
	746.438468873032	0.994027409604575\\
	746.389979084646	0.99402350783234\\
	746.365449680118	0.994019638502628\\
	746.321850393701	0.994015802486293\\
	746.282203494094	0.994011976805617\\
	746.248746616634	0.994008176763493\\
	746.211614173228	0.99400440113778\\
	746.186123585138	0.994000654892047\\
	746.14864511565	0.993996939142233\\
	746.111366572342	0.993993239699652\\
	746.085191621555	0.993989561846442\\
	746.050058439961	0.993985903590111\\
	746.034187376968	0.993982264825224\\
	745.997393270177	0.993978645834714\\
};

\addplot [color=mycolor6,solid,mark=+,mark options={solid,scale=3,thick}]
table[row sep=crcr]{%
	738.526859313484	0.989667951729153\\
};

\addplot [color=mycolor2,solid,mark=square,mark options={solid},mark repeat = 11,mark phase = 1]
table[row sep=crcr]{%
	757.381020853839	0.994465214269777\\
	756.974263348917	0.994459843937192\\
	756.646599717028	0.994453854791302\\
	756.364350086122	0.994447613457406\\
	756.125507504921	0.994441248213124\\
	755.936546505906	0.994434882050297\\
	755.761634165846	0.99442858327896\\
	755.605168860728	0.994422411056265\\
	755.463967150591	0.994416389439578\\
	755.335968257874	0.994410531043522\\
	755.226739357776	0.99440484879962\\
	755.122485543799	0.994399313846126\\
	755.035594549705	0.994393976063621\\
	754.945758489173	0.994388828851233\\
	754.8642578125	0.994383860020564\\
	754.790800319882	0.994379066250133\\
	754.710622231791	0.994374433172342\\
	754.635288508858	0.994369966523482\\
	754.579747477854	0.994365655553565\\
	754.528820127953	0.99436150279487\\
	754.456039308563	0.994357482801933\\
	754.406280757874	0.994353606354453\\
	754.358959768701	0.994349849413574\\
	754.300696665846	0.994346237342965\\
	754.261111281988	0.994342745581163\\
	754.208384596457	0.994339384612056\\
	754.174697034941	0.994336119894137\\
	754.121878075787	0.994332974810231\\
	754.085583784449	0.99432993323479\\
	754.035440760335	0.994327000471203\\
	754.016094057579	0.99432416550166\\
	753.977992741142	0.994321418667828\\
	753.947603961614	0.9943187670328\\
	753.909625676673	0.99431618648482\\
	753.884081262303	0.994313695641415\\
	753.862842950295	0.994311275543677\\
	753.827340674213	0.994308932544744\\
	753.792868786909	0.994306670952498\\
	753.76624015748	0.994304463199966\\
	753.734959399606	0.994302330982134\\
	753.719096026083	0.994300275377313\\
	753.686861774114	0.994298266911423\\
	753.664500799705	0.994296312394732\\
	753.632558747539	0.994294419311077\\
	753.602992741142	0.994292576381551\\
	753.587598425197	0.994290798288466\\
	753.560177780512	0.994289063874535\\
	753.542476624016	0.994287385997127\\
	753.52307609498	0.994285749094204\\
	753.497270238681	0.994284161465986\\
	753.482721764272	0.994282620964213\\
	753.456731360728	0.994281116081638\\
	753.430110420768	0.994279654058433\\
	753.426273375984	0.994278224251329\\
	753.397084153543	0.994276843315805\\
	753.382304995079	0.99427549820349\\
	753.37186269685	0.994274189234061\\
	753.34884042815	0.994272913518679\\
	753.337213951772	0.99427167168338\\
	753.326248769685	0.994270451394896\\
	753.303787832185	0.994269273266211\\
	753.287155511811	0.994268118406209\\
	753.280950110728	0.994266992307203\\
	753.258450725886	0.994265898024792\\
	753.243194820374	0.994264831818046\\
	753.243586983268	0.994263793358419\\
	753.214751476378	0.994262770569946\\
	753.206231545276	0.994261777487618\\
	753.198426734744	0.994260806699207\\
	753.178949311024	0.994259864849264\\
	753.162578432579	0.994258939947266\\
	753.166653850886	0.994258044969986\\
	753.146968811516	0.994257165154087\\
	753.140025221457	0.994256309099809\\
	753.13003660187	0.994255473458705\\
	753.123346764272	0.994254653224188\\
	753.097871555118	0.99425386278745\\
	753.096702755906	0.994253082595757\\
	753.086368110236	0.994252312702292\\
	753.074787770669	0.994251563554388\\
	753.07092765748	0.994250831830787\\
	753.05627921998	0.994250105862333\\
	753.043445497047	0.994249401276817\\
	753.037109375	0.994248719927416\\
	753.01608636811	0.994248050335491\\
	753.01593257874	0.994247396105833\\
	753.010811392717	0.994246755096858\\
	752.992156742126	0.994246130484117\\
	752.983106237697	0.994245517677706\\
	752.974901574803	0.994244910380437\\
	752.965528112697	0.994244321717921\\
	752.966758427658	0.994243745110502\\
	752.952886626476	0.994243182929932\\
	752.940421998032	0.994242626689398\\
	752.927026943898	0.994242087175022\\
	752.925181471457	0.994241550482817\\
	752.92022945374	0.994241024757257\\
	752.912140132874	0.994240510848236\\
	752.898929625984	0.994240013267192\\
	752.899644746555	0.994239521707304\\
};

\addplot [color=mycolor1,solid,mark=asterisk,mark options={solid,scale=3,thick}]
table[row sep=crcr]{%
	758.434685654528	0.994472191585476\\
};

\end{groupplot}


\path (group c1r1.south west |-current bounding box.south west)--
coordinate(legendpos)
(group c2r1.south east |-current bounding box.south east);
\matrix[
matrix of nodes,
anchor=south,
inner sep=0.2em,
draw,
nodes={scale=0.9},
]at([yshift=-6ex,xshift = -0ex]legendpos)
{
	\ref{none}& MCTF&[5pt]
	\ref{gaussian} & $\text{WLDU}_\text{Gauss}$ &[5pt]
	\ref{AWF}& $\text{WLDU}_\text{AWF}$ &[5pt]
	\ref{NLM}& $\text{WLDU}_\text{NLM}$ & &[5pt]
	\ref{BM3D}& $\text{WLDU}_\text{BM3D}$ & &[5pt]
	\ref{GIF}& $\text{WLDU}_\text{GIF}$ & &[5pt]
	\ref{trunc} & Truncated WT &[5pt]\\};


\draw[thick] (2.84,2.99) circle (3.5pt);
\draw[thick] (2.15,1.85)  circle (3.5pt);
\draw[thick] (3.945,3.03) circle (3.5pt);
\draw[thick] (0.53,0.35) circle (3.5pt);

\end{tikzpicture}%